\documentclass[11pt]{article}
\newcommand{\be}{\begin{eqnarray}}\newcommand{\beq}{\begin{equation}}
\newcommand{\ee}{\end{eqnarray}}\newcommand{\eeq}{\end{equation}}
\newcommand{\ep}{\epsilon}
\newcommand{\la}{\lambda}\newcommand{\De}{\Delta}
\newcommand{\rad}{\raisebox{1.3ex}{\tiny{\ensuremath{\bullet}}}}
\usepackage{graphicx,amssymb,amsmath} 
\title{Kinetic equation of concurrent nucleation and chemical aging 
of an ensemble of 
aqueous organic aerosols  
}
\author{ 
Yuri S. Djikaev\thanks{Corresponding author. E-mail: idjikaev@buffalo.edu} 
\hspace{0.1cm} and \hspace{0.1cm}Eli Ruckenstein\thanks{ E-mail: feaeliru@buffalo.edu}\hspace{0.1cm}
\\ Department of Chemical and Biological  Engineering, SUNY at Buffalo, \\ Buffalo, New York  14260 
}
\date{}
\renewcommand{\baselinestretch}{1}
\begin{document}
\bibliographystyle{plain}
\renewcommand{\baselinestretch}{1}
\maketitle
\begin{abstract}
{\bf \LARGE }

Using the formalism of the classical nucleation theory,   we derive a novel kinetic equation for the 
size and composition distribution of an ensemble of aqueous organic aerosols, evolving 
via nucleation and concomitant chemical aging.  
This distribution can be drastically affected by the enthalpy of heterogeneous chemical reactions and
the depletion of organic trace gases absorbed by aerosols. 
A partial differential equation of second order for the
temporal evolution of this distribution is obtained from the discrete equation of balance via Taylor
series expansions.  Once reduced to the canonical form of the multidimensional Fokker-Planck equation,
this kinetic equation can be solved via the method of  complete separation of variables. 
The new kinetic equation opens a new direction in the development of the kinetic theory of
first-order phase transitions, while its applications to the formation and evolution atmospheric organic
aerosols may drastically improve the accuracy of global climate models.

\end{abstract}

\begin{figure}[h]
\begin{center}\vspace{0.3cm}
\vspace{-3.2cm}
\end{center}
\end{figure} 
\renewcommand{\baselinestretch}{1} 
\renewcommand{\baselinestretch}{2}
\newpage
\normalsize
\renewcommand{\baselinestretch}{2}

Liquid organic aerosols (OAs)   constitute a significant fraction  of atmospheric particulate matter
[1-3].$^{ }$ They have a  high impact on the Earth climate, air quality, and human health by directly
contributing to both the scattering and adsorption of solar radiation [4-6].$^{}$  Particularly widespread
are aqueous aerosols consisting of water and both hydrophobic and hydrophilic compounds [2,3].$^{}$
Hereafter, such an aerosol particle will be referred to as an aqueous hydrophilic/hydrophobic organic
(AHHO) aerosol. 

Although the chemical composition of OAs can be extremely  complex [2,3],$^{}$ one can anticipate that the
hydrophilic parts of   organic compounds of an aqueous OA particle will be embedded into its aqueous
core,  leaving the hydrophobic parts pointed outward [7].$^{}$ Hydrophobic compounds at the surface of an
aerosol hinder its ability to act as  a cloud condensation nucleus [8].$^{}$ However, the hygroscopicity
of such aerosols can significantly increase via several mechanisms [9-11],$^{}$ of which the ``chemical
aging" is the most  widespread and least understood.$^{}$ This mechanism consists of  heterogeneous
chemical reactions between aerosol surface-located hydrophobic molecules and  some atmospheric gaseous
species which enable aerosols to  become cloud condensation nuclei (CCN) [7].$^{}$ 

In the atmosphere, the chemical aging of OAs occurs concomitantly [7]$^{}$ with the   condensation of
water and other vapors, such as oxidized volatile and semi-volatile organic compounds (OVOCs and SVOCs,
respectively).  Recently, in the framework of classical nucleation theory (CNT)  we developed a
thermodynamic model [12-16]$^{}$ for the  formation and evolution of aqueous organic aerosols via
concurrent nucleation/condensation and chemical aging. 

That model allows one  to determine the threshold parameters necessary for the K\"ohler activation of such
aerosols [12-16].$^{}$  We have also shown that the size and composition distribution of an {\em ensemble}
of aqueous organic aerosols, evolving via nucleation/condensation and concomitant chemical aging, may be
affected by the release of the enthalpy of heterogeneous reactions and by the depletion of atmospheric
{\em trace}  gases [17],$^{}$ OVOCs, and SVOCs (hence decrease of their saturation ratios) due to their
uptake by aerosols.  To properly account of these effects in global climate models,  it is necessary to
develop an analytical kinetic model for the temporal evolution  of an ensemble of atmospheric aqueous
organic aerosols occurring via both condensation and  concomitant chemical aging. 

As a foundation for such a theory, we hereafter present a kinetic equation for    the size and composition
distribution function of an ensemble of AHHO aerosols  evolving via concurrent nucleation and chemical aging.
We develop our model by using the formalism of the  CNT and treating a particle of a new phase (liquid
aerosol) in the framework of capillarity approximation [18,19].$^{}$ 

Consider an air parcel containing a ternary mixture of condensable vapors -- water and  hydrophilic and
hydrophobic organics (components 1, 2, and 3, respectively), as well as non-condensable species -- nitrogen
oxide,  hydroxyl radicals, oxygen, and nitrogen dioxide; other air components are neglected.   An aerosol
itself is modeled as a spherical  particle of a liquid multicomponent solution. Initially, it contains
components 1, 2, and 3  due to their condensation  from the surrounding air.   Being hydrophobic, component 3
is mostly (but possibly not exclusively, if it also contains a hydrophilic moiety, however weak) located on
the aerosol surface, forming  hydrophobic patches there. However, owing to chemical reactions with
atmospheric species,   molecules of component 3 can be transformed into hydrophilic entities. 

Such a ``hydrophobic-to-hydrophilic" conversion  (i.e., chemical aging of organic aerosols) may proceed via
diverse pathways each involving a variety of sequential heterogeneous reactions.   However,  it is most
likely initiated [7]$^{}$   by atmospheric OH radicals abstracting H-atoms from the  hydrophobic moieties of
surfactant molecules on the aerosol surface. 

Denote a hydrophobic/surfactant molecule by HR, with the radical  ``R\hspace{-0.02cm}{\rad}"  being
the entire  molecule less one of the hydrogen atoms, ``H", in its hydrophobic moiety.
The first three steps of the most probable sequence of reactions,
involved in the chemical mechanism of aerosol aging, are [7]:$^{}$ 
\beq \mbox{OH (g)+ HR/aerosol}\rightleftharpoons 
\mbox{H$_2$O (g) + R\hspace{-0.02cm}{\rad}/aerosol}.  \eeq
\beq \mbox{O$_2$ (g) + R\hspace{-0.02cm}{\rad}/aerosol}\rightleftharpoons
\mbox{RO$_2$\hspace{-0.16cm}{\rad}/aerosol}. \eeq
\beq 
\mbox{RO$_2$\hspace{-0.16cm}{\rad}/aerosol+NO (g)}
\rightleftharpoons\mbox{RO\hspace{-0.02cm}{\rad}/aerosol+NO$_2$ (g) }. \eeq 

On the first step of this pathway, an OH radical abstracts an H atom from the  hydrophobic moiety of a
surfactant molecule, thus producing a surface-bound radical R\hspace{-0.02cm}{\rad} (reaction (1)). The
latter  is rapidly oxidized by O$_2$ molecules at the second step of the sequence, thus producing a new
surface-bound radical RO$_2$\hspace{-0.16cm}{\rad} (reaction (2)). The rate constants of reactions (1) and (2)
are of the same order [7],$^{}$  but the concentration of oxygen molecules in the atmosphere is much higher (by
$\sim$12 orders of magnitude) than that of OH radicals. Hence, every radical R\hspace{-0.02cm}{\rad}  produced
by reaction (1) is almost immediately oxidized by reaction (2). 
The further evolution
of radicals RO$_2$\hspace{-0.16cm}{\rad} may  vary, but always
results in the formation of  water soluble and/or volatile species and/or hydrophilic radicals [7].$^{}$  
Reaction (3) represents one such a pathway 
(see ref.7 for the discussion of various reactive channels of radicals
RO$_2$\hspace{-0.16cm}{\rad} and RO\hspace{-0.01cm}{\rad}).

\par As a result of sequential reactions (1)-(3), a surface located hydrophobic molecule HR is converted 
into a radical RO\hspace{-0.02cm}{\rad}. The latter may still contain hydrophobic parts, but there
now appears at least one highly hydrophilic site on its formerly hydrophobic moiety. Consequently,   
radicals  RO\hspace{-0.02cm}{\rad} will be able to diffuse into the aerosol interior. 
According to recent numerical evaluations  [14],$^{}$ the characteristic time of the entire
sequence of  reactions (1)-(3) is significantly shorter than the characteristic time of the evolution of the
total number of molecules in the aerosol. Therefore, the number of intermediate radicals
R\hspace{-0.01cm}{\rad} 
and RO$_2$\hspace{-0.1cm}{\rad} 
in
the aerosol can be assumed negligible compared to the  number of ``final" radicals RO\hspace{-0.01cm}{\rad}, 
so that the entire sequence  (1)-(3) produces only one additional component in the
aerosol, namely, radical  RO\hspace{-0.01cm}{\rad}; it will be referred to as component 4. 

Denote the numbers of
molecules of components $1$ (water), $2$ (hydrophilic organic), and $3$ (hydrophobic organic) in the aerosol
will be denoted by $\nu_1,\nu_2$, and $\nu_3$, respectively, and the number of radicals 
RO\hspace{-0.01cm}{\rad} (component 4) in the aerosol  by $\nu_4$. For the sake of simplicity
and uniformity, the radicals RO\hspace{-0.01cm}{\rad} will be also referred to as ``molecules of
component $4$".  

Choosing $\nu_1,\nu_2,\nu_3$, and $\nu_4$ as the independent  variables of state of a single aerosol, consider
an ensemble of AHHO aerosols and  denote their distribution function with   respect to $\nu_1,\nu_2,\nu_3$,
and $\nu_4$ at time $t$ by $g(\nu_1,\nu_2,\nu_3,\nu_4,t)$.   Hereafter, for the sake of convenience, 
any function $f$ of variables  $\nu_1,\nu_2,\nu_3,\nu_4$ may be denoted either 
$f(\nu_1,\nu_2,\nu_3,\nu_4)$ or  $f(\{\nu\})$ or $f(\nu_i,\widetilde{\nu_i})$, where $\{\nu\}$ 
would represent all four variables $\nu_1,..,\nu_4$, whereas the ``complementary"
variable  $\widetilde{\nu_i}$ represents only three of them,  the ``excluded" variable being $\nu_i$. 
Thus, e.g., $g(\{\nu\},t)=g(\nu_i,\widetilde{\nu_i},t)=g(\nu_1,\nu_2,\nu_3,\nu_4,t)$.

Let us construct a discrete balance equation governing the temporal evolution of 
$g(\nu_1,\nu_2,\nu_3,\nu_4,t)$. 
As usual in CNT,$^{18-20}$ assume that the metastability of vapor mixture in the air is  created
instantaneously,  
its temperature $T$ and the number densities of
non-condensable  gas molecules are also constant, and 
every aerosol attains its internal thermodynamical equilibrium 
before each successive interaction with air, so that  its temperature is equal to the air temperature $T$.

\par The material exchange between an aerosol and air occurs via three types of elementary interactions: 
1) absorption of a molecule of component $i\;\;(i=1,2,3)$ from the air into the aerosol; 
2) emission of a molecule of component $i\;\;(i=1,2,3)$ from the aerosol into the air; 
3) sequence (1)-(3) of forward and backward chemical reactions on the aerosol surface whereby a 
radical  RO\hspace{-0.01cm}{\rad} is either produced or destroyed. 
Therefore, the temporal evolution of the distribution
function $g(\{\nu\},t)$ can be described by the discrete balance equation [20]  
\beq \frac{\partial g(\{\nu\},t))}{\partial t}=
-\sum_{i=1}^{3}[J_i(\nu_i+1,\widetilde{\nu_i},t)-J_i(\{\nu\},t)]-[J_4(\nu_1,\nu_2-1,\nu_3,\nu_4+1,t)-
J_4(\{\nu\},t)],
\eeq
where the fluxes $J_i(\nu,t)\;\;(i=1,..,4)$ 
are defined as 
\beq J_i(\nu,t)= W^+_i(\nu_i-1,\widetilde{\nu_i})g(\nu_i-1,\widetilde{\nu_i},t)-
W_i^-(\{\nu\})g(\{\nu\},t)\;\;\;\;\;\;\;(i=1,2,3),\eeq
\beq J_4(\{\nu\},t)=W_4^+(\nu_1,\nu_2-1,\nu_3,\nu_4+1)g(\nu_1,\nu_2-1,\nu_3,\nu_4+1)-
W_4^-(\{\nu\})g(\{\nu\},t),\eeq 
with $W_i^+=W_i^+(\{\nu\})\;\;(i=1,2,3$ and $W^-_i= W_i^-(\{\nu\})\;\;(i=1,2,3)$ being 
the numbers of molecules of component $i$ that the aerosol absorbs from and emits into the air (``absorption"
and ``emission" rates, respectively), 
and $W_4^+=W_4^+(\{\nu\})$ and $W_4^-=W_4^+(\{\nu\})$ 
the aggregate rates of forward and backward reactions, respectively, of the sequence (1)-(3) 
(the evolution of aerosols is assumed to occur through the absorption from and emission into
the vapor of single molecules of components $1,2$, and $3$, or through a single sequences (1)-(3)
of forward and backward reactions whereby a radical  RO\hspace{-0.01cm}{\rad} is either formed or destroyed). 

In CNT, the absorption rate $W_i^+\;\;(i=1,2,3)$ is provided by
the gas-kinetic theory, whereas the emission rate $W_i^-$ is determined
through  $W_i^+$ by using the principle of detailed
balance [18,19].$^{}$  
It requires that for the equilibrium distribution function $g_{\mbox{\tiny
e}}(\{\nu\})$ 
every equilibrium flux $J_i^{\mbox{\tiny e}}\equiv J_i^{\mbox{\tiny e}}(\{\nu\})\;\;(i=1,..,4)$
be equal to 0. Applying this principle to $J_4^{\mbox{\tiny e}}$,  one can obtain [20] an interesting
relationship between the aggregate rates of forward and backward sequence of heterogeneous chemical reactions 
(1)-(3),
\beq 
W_4^-(\{\nu\})\simeq W_4^+(\{\nu\})[1-(F'_2(\{\nu\})-F'_4(\{\nu\}))],\eeq
where $F(\{\nu\})$ denotes the free energy of formation of an AHHO aerosol (expressed in thermal units
$k_BT$, with $k_B$ being Boltzmann's constant), and  $F'_i(\nu)=\partial F(\nu)/\partial
\nu_i\;\;(i=1,..,4)$. 
Clearly, this relationship 
can be also used to obtain a relationship between the corresponding reaction rate constants. 

During nucleation, aerosols overcome a free-energy barrier (5D surface 
determined by the function $F=F(\{\nu\})$) 
to become 
growing droplets. 
At this stage, aerosols evolve in the near-critical region of the  space of variables
$\nu_1,..,\nu_4$, i.e., in the vicinity of the ``saddle point" of the  free-energy surface [13]$^{}$ 
with the coordinates
$\nu_{1c},\nu_{2c},\nu_{3c},\nu_{4c}$ (subscript ``c" marks quantities for the saddle point). 
Assuming that in this vicinity $\nu_i\gg 1\;\;(i=1,..,4)$, expanding
eqs.(4)-(6) in Taylor series in the deviations of aerosol   characteristics from $\nu_1,\nu_2,\nu_3$, and
$\nu_4$ and retaining only leading terms therein, and with the same accuracy approximating 
$W_{i}^+\approx W_{ic}^+\;\;(i=1,..,4)$,   one can [20] reduce eq.(4) 
to 
\be \frac{\partial g(\{\nu\},t))}{\partial t}&=&\sum_{i=1}^{4}
W_{ic}^+\frac{\partial }{\partial \nu_i}
\left( F'_i(\{\nu\})+\frac{\partial }{\partial \nu_i} \right) g(\{\nu\},t))\nonumber \\
&&+W_{4c}^+
\left[ \frac{\partial }{\partial \nu_2}
\left(F'_2(\{\nu\})+\frac{\partial }{\partial \nu_2}\right)-
\frac{\partial }{\partial \nu_2}
\left(F'_4(\{\nu\})+\frac{\partial }{\partial \nu_4}\right)-
\frac{\partial }{\partial \nu_4}\left(F'_2(\{\nu\})+\frac{\partial }{\partial \nu_2}\right)
\right]g(\{\nu\},t)).
\ee
This is the kinetic equation of concurrent multicomponent nucleation and chemical
aging of an ensemble of model organic aerosols. 
Note that the second term on the RHS of eq.(8) arises because any sequence of reactions (1)-(3) is
results in the change of not only $\nu_4$, but also $\nu_2$; the latter also changes
independently via the direct exchange of component 2 between aerosol and air. 

By definition of the near-critical region [18,21],$^{}$ 
$F(\{\nu\})$ can be approximated 
there as a bilinear form 
\beq F(\{\nu\})=F_c+\frac1{2}\sum_{i=1}^4F''_{ijc}(\nu_i-\nu_{ic})(\nu_j-\nu_{jc})=F_c+
{\bf \De\nu}^{\mbox{\tiny T}}{\bf A}{\bf \De\nu}, 
\eeq 
where $F''_{ij}=\partial^2 F/\partial\nu_i \partial\nu_j\;\;(i=1,..,4)$   
and  the matrix notation 
was introduced 
with a real symmetric  $4\times 4$-matrix ${\bf A}=[a_{ij}]\equiv [\frac1{2}F''_{ijc}]\;\;(i,j=1,..,4)$ 
and a  real column-vector of length 4  ${\bf \De\nu}=[\De\nu_i]\equiv [\nu_i-\nu_{ic}]\;\;(i=1,..,4)$;  
superscript ``$\mbox{T}$" marks 
the transpose of a matrix/vector.   In this approximation, 
the first derivatives $F'_i(\{\nu\})\;\;(i=1,..,4)$  are linear
superpositions of deviations $\De\nu_i \;\;(i=1,..,4)$. Thus,  the accuracy of eq.(8) combined with eq.(9)
corresponds to the Fokker-Planck approximation [22],$^{}$ widely used in the kinetics of the
first-order phase transitions [18,19,21].$^{}$

Equation (8) must be solved subject to appropriate boundary conditions. In the framework of CNT, 
the latter are formulated by using the equilibrium distribution $g_e(\{\nu\})$ [18,21].$^{}$  
In the near-critical 
region, $g_e(\{\nu\})$ can be written (in virtue of eq.(9)) as  [21]$^{}$ 
\beq g_e({\bf \nu})= 
(1/v_c)\exp\left[-F_c-\frac1{2}\sum_{i=1}^4F''_{ijc}\De\nu_i\De\nu_j\right],\eeq 
where $v_c$ is the volume per molecule in the saddle point aerosol (nucleus). 
As clear, the variables $\nu_1,..,\nu_4$ are not separated in the equilibrium distribution (nor in the
kinetic equation (8)). Hence, they are not convenient to formulate simple enough boundary 
conditions to the kinetic equation (8). 

The problem can be overcome by using an elegant method of complete separation of variables  developed by
Kuni and co-workers [23,21].$^{}$  Its most general presentation can be found in Ref.[24]. 

First, it is necessary to introduce such new variables (instead of $\nu_1,..,\nu_4$) that would transform the
bilinear form in eq.(9) into a quadratic one, without cross terms. 
Let us denote a set of such variables by $\{ x\}=(x_1,..,x_4)$ and the corresponding column-vector by 
${\bf x}=[x_i]\;\;(i=1,..,4)$. 

According to the spectral theorem [25],$^{}$ matrix ${\bf A}$ is orthogonally diagonalizable
because it is real and symmetric. Hence, there exists a real orthogonal 
$4\times 4$-matrix  ${\bf P}\equiv [p_{ij}]\;\;(i,j=1,..,4)$, such that 
\beq {\bf P}^{\mbox{\tiny T}}{\bf A}{\bf P}={\bf D},\eeq
where 
the columns of the matrix ${\bf P}$ are $4$ linearly independent orthonormal eigenvectors of  ${\bf A}$,  
and the diagonal elements of the diagonal matrix ${\bf D}$ are the corresponding 
eigenvalues $\lambda_1,..,\lambda_4$ of ${\bf A}$. 

Under conditions when a multicomponent first-order phase transition occurs via nucleation, the corresponding
free energy surface has a shape of a hyperbolic paraboloid (``saddle" shape in three dimensions). Therefore,
one of the eigenvalues of matrix ${\bf A}$ is negative while all others are positive, so that $\det({\bf
A})<0$. Denoting the negative eigenvalue by $\lambda_1$, one can thus define the variables
\beq  x_i=|\lambda_i|^{1/2}\sum_{j=1}^4p_{ji}\De\nu_j\;\;\;(i=1,..,4),\eeq 
in which the bilinear form (9) transforms into  
\beq F(\{x\})=F_c-\sum_{i=1}^4\epsilon_ix_i^2\;\;\;(\epsilon_1=1,\;\epsilon_2=\ep_3=\ep_4=-1).\eeq 

The transformation of variables (12) allows one to re-write the kinetic equation of nucleation
and chemical aging, eq.(8), in the canonical form of Fokker-Planck equation
\beq 
\frac{\partial g(\{x\},t)}{\partial t}=
\sum_{i,j=1}^{4}b_{ij}\frac{\partial }{\partial x_i}(\frac{\partial }{\partial
x_j}-\epsilon_j2x_j)]g(\{x\},t),
\eeq
where 
$b_{ij}\;\;(i,j=1,..,4)$ are the elements of a
$4\times 4$ matrix ${\bf B}$ of diffusion coefficients in variables $\{x\}$,  
\beq b_{ij}=|\la_i\la_j|^{1/2}\left[\sum_{k=1}^4W_{kc}^+p_{ki}p_{kj}-W_{4c}^+(p_{2i}p_{2j}-p_{2i}p_{4j}-p_{4i}p_{2j})\right].\eeq
Clearly, the matrix ${\bf B}$ is real and symmetric, but its positiveness is not obvious. 
The physically reasonable 
boundary conditions to equation (14) can be now 
rigorously formulated: 
\beq \frac{g(\{x\},t)}{g_e(\{x\})}=\left\{\begin{array}{ll} 1 & \;\;(x_0\rightarrow -\infty\;\;
\mbox{and}\;\;\sum_{i=1}^4\epsilon_ix_i^2>0),\\
0 & \;\;(x_0\rightarrow \infty\;\;
\mbox{and}\;\;\sum_{i=1}^4\epsilon_ix_i^2>0), \end{array} \right. \eeq
and $g(\{x\},t)\rightarrow 0$ as $\sum_{i=1}^4\epsilon_ix_i^2\rightarrow -\infty$. 

Solving the kinetic equation (14) subject to conditions (16) still remains a
formidable task because the variables $x_1,..,x_4$ do not separate in eq.(14) - the diffusion 
matrix $B$ is not diagonal. Such a difficulty can be overcome by introducing new variables $\xi_1,..,\xi_4$
via the Lorentz transformation, whereof the Jacobian equals 1 and which leaves 
the quadratic form $\sum_{i}\epsilon_ix_i^2$ invariant,
so that 
\beq \xi_i=\sum_{j}L_{ij}x_j,\;\;\;F=F_c-\sum_{i=1}^4\epsilon_i\xi_i^2,\;\;\;
g(\{x\},t)=g(\{\xi\},t), \eeq 
where $L_{ij}$ are the Lorentz transformation coefficients.  
This  allows one to find the solution of the kinetic equation and thoroughly
investigate its properties. The most general formulation of the method is presented in Ref.[24], while its
applications to various particular problems of the kinetics of first-order phase transitions {\em without
chemical reactions} were reported in Ref.[23].

As a numerical illustration of the foregoing, we considered the concurrent nucleation and chemical aging of 
model AHHO aerosols 
in the air  containing a ternary vapor mixture of water, $2-$methylglyceric acid
[C$_4$H$_8$O$_4$], 
and $3-$methyl$-4-$hydroxy-benzoic acid [C$_8$H$_8$O$_3$],
as well as nitrogen oxide, hydroxyl radicals, oxygen, and nitrogen dioxide 
According to Couvidat {\em et al.} [17],$^{}$ $2-$methylglyceric acid can be
classified as a hydrophilic compound, whereas  $3-$methyl-4-hydroxy-benzoic acid can be characterized as a
hydrophobic one. The atmospheric conditions were 
assumed to be as in Refs.[13-16].

Hydrophobic molecules of $3-$methyl-4-hydroxy-benzoic acid$^{}$ are mostly 
located at the aerosol surface, with the methyl groups -CH$_3$ 
exposed to the air. 
The abstraction of an H-atom from the methyl group of a $3-$methyl-4-hydroxy-benzoic acid molecule 
can be considered as the first step in its hydrophobic-to-hydrophilic conversion, so that 
the radicals R\hspace{-0.01cm}{\rad} and RO\hspace{-0.01cm}{\rad} 
in eqs.(1)-(3) are 
can be identified as 
$ \mbox{-CH$_2$-C$_6$H$_3$-OH-COOH}$ and $\mbox{-OCH$_2$-C$_6$H$_3$-OH-COOH}$, 
respectively.
Thus, one can consider the 4-component solution in the aerosol as a liquid mixture of functional
groups whereof all relevant parameters for 
activity coefficients are given in the updated tables of UNIFAC method [13].$^{}$ 

The free energy of formation of an AHHO aerosol as a function of four variables
$F=F(\nu_1,..,\nu_4)$ was obtained in Ref.[13]. 
Its second derivatives at the saddle point (under the same conditions as in Refs.[13-16])  
provide the matrix ${\bf A}$ in eq.(9). This matrix ${\bf A}$, the corresponding 
orthogonal matrix ${\bf P}$ which diagonalizes ${\bf A}$, and the resulting diagonal matrix 
${\bf D}={\bf P}^T{\bf A}{\bf P}$ are 
\be  
{\bf A}&=&
   \begin{bmatrix}
  
    0.00508819  & -0.0102213  &  0.000425629 & -0.00779759 \\
   -0.0102213   &  0.029559   & -0.00841285  & -0.00106402 \\
    0.000425629 & -0.00841285 &  0.0411663   & -0.0144458  \\
   -0.00779759  & -0.00106402 & -0.0144458   &  0.0108751   \\
   \end{bmatrix}, \nonumber \\
{\bf P}&=&
   \begin{bmatrix}
   -0.151706 &  0.283741 &  0.60688  & 0.726755 \\
    0.393328 & -0.859266 &  0.155576 & 0.287666 \\
   -0.846176 & -0.367706 & -0.311695 & 0.227209 \\
    0.325981 &  0.214351 & -0.71438  & 0.580906 
     \end{bmatrix}, \nonumber \\
{\bf D}&=& 
   \begin{bmatrix}
   0.0507183 & 0    	 & 0         & 0           \\
   0         & 0.0295995 & 0         & 0           \\
   0         & 0         & 0.0114281 & 0           \\
   0         & 0         & 0         & -0.00505728 
   \end{bmatrix}\nonumber 
\ee 
As required, 
the diagonal elements $\lambda_1,..,\la_4$ of the matrix ${\bf D}$ and the corresponding columns
$\overline{\bf P}_1,..,\overline{\bf P}_4$ of the orthogonal matrix ${\bf P}$ represent the 
eigenvalues and eigenvectors of the original  
matrix ${\bf A}$, so that ${\bf A}\overline{\bf P}_i=\lambda_i\overline{\bf P}_i\;\;(i=1,..,4)$.
(Numerical orthogonal diagonalization was performed with Mathematica 11.3; the 
off-diagonal elements of ${\bf D}$ are equal to zero with the accuracy of $10^{-16}$ or better).

In summary, we have derived the kinetic equation for the distribution function of an ensemble of aqueous 
organic aerosols evolving via concurrent nucleation and chemical aging. The proposed equation takes account of
heterogeneous chemical reactions on the surface of aerosols and hence differs from the classical kinetic
equation of multicomponent nucleation. The principle of detailed balance  allows one to obtain a relationship
between aggregate forward and backward  rates (and hence rate constants) of the sequence of reaction whereby
the chemical aging occurs.  This relationship can be experimentally verified in appropriate experiments  or it
can serve as a means for the indirect experimental  determination of the rates or rate constants of some of
the chemical reactions involved.   We have also reduced the new kinetic equation to  the canonical form of a
4-dimensional Fokker-Planck equation, whereof the diffusion coefficients depend on the aggregate rate of the
forward sequence of chemical reactions, and formulated rigorous  boundary conditions to this equation. As
outlined, this kinetic  equation can be solved by using the  method of complete separation of variables
developed in CNT  for the kinetics of multidimensional first-order phase transitions {\em without}   chemical
reactions.

\section*{References}
\begin{list}{}{\labelwidth 0cm \itemindent-\leftmargin}
\item $[1]$ United States Environmental Protection Agency, {\it Air quality criteria for
particulate matter, EPA/600/P-95/001} (Environ. Prot. Agency, Washington, D.C., 1996).
\item $[2]$ P. Saxena and L. M. Hildemann,  
J. Atmos. Chem.  {\bf  24}, 57-109 (1996).
\item $[3]$ M. Kanakidou, J. H. Seinfeld, S. N. Pandis, I. Barnes, F. J. Dentener, M. C. Facchini,  
R. Van Dingenen, B. Ervens, A. Nenes, C.J. Nielsen, et al. 
Atmos. Chem. Phys.  {\bf  5}, 1053-1123 (2005).
\item $[4]$ J. H. Seinfeld and S. N. Pandis,  
{\it Atmospheric chemistry and physics: from air pollution to climate
change} (John Wiley \& Sons, New York, 2006).
\item $[5]$ C. A. Pope,  
J.Aerosol Med.  {\bf  13}, 335-354 (2000).
\item $[6]$ B. J. Finlayson-Pitts and J. N. Pitts, Jr. {\it Atmospheric chemistry:
fundamentals and experimental techniques}  (John wiley \& Sons, New 
York, 1986). 
\item $[7]$ G. B. Ellison, A. F. Tuck, and V. Vaida, 
J. Geophys. Res.  {\bf  104}, 11633-11641 (1999).
\item $[8]$ C. R. Ruehl and K. R. Wilson,  
J. Phys. Chem. A.  {\bf  118}, 3952-3966 (2014).
\item $[9]$ M. D. Petters, A. J. Prenni, S. M. Kreidenweis, P. J. DeMott, A. Matsunaga, Y. B. Lim, and 
P. J. Ziemann,  
Geophys. Res. Lett.  {\bf  33},  L24806 (2006).
\item $[10]$ Y. Rudich,  
Chem. Rev. {\bf  103}, 5097-5124 (2003).
\item $[11]$  Y. Rudich, N. M. Donahue, and T. F. Mentel, 
Annu. Rev. Phys. Chem. {\bf  58}, 321-352 (2007).
\item $[12]$ Y. S. Djikaev and E. Ruckenstein, 
J. Phys. Chem. A {\bf  118}, 9879-9889 (2014).
\item $[13]$ Y. S. Djikaev and E. Ruckenstein, 
J. Phys. Chem. A {\bf  122}, 4322-4337 (2018).
\item $[14]$ Y. S. Djikaev and E. Ruckenstein, 
J. Phys. Chem. Lett. {\bf  9}, 5311-5316 (2018).
\item $[15]$ Y. S. Djikaev and E. Ruckenstein,  
Adv. Colloid Interface Sci. {\bf  265}, 45-67 (2019).
\item $[16]$ Y. S. Djikaev and E. Ruckenstein,  
Phys. Chem. Chem. Phys. {\bf  265}, 13090-13098 (2019). 
\item $[17]$ F. Couvidat, E. Debry, K. Sartelet and C. Seigneur, 
J. Geophys. Res. {\bf  117}, D10304 (2012). 
\item $[18]$ D. Kaschiev, {\it Nucleation : basic theory with applications}  
(Butterworth Heinemann, Oxford, Boston, 2000). 
\item $[19]$ E. Ruckenstein and G. Berim, {\it Kinetic theory of nucleation} (CRC: New York, 2016).
\item $[20]$ See   Supplemental   Material   at www.... for more details of 
the derivation of equations (4), (7), and (8). 
\item $[21]$ A. A. Melikhov, V. B. Kurasov, Y. S. Dzhikaev, and F. M. Kuni, 
Khim. Fiz. (in Russian) {\bf 9}(12), 1713-1722 (1990); 
Sov. Phys. Tech. Phys. {\bf 36}, 14-19 (1991); 
Zh. Tekh. Fiz. (in Russian) {\bf 61}(1), 27-34 (1991).
\item $[22]$ H. Kramers, Physica (Utrecht) {\bf 7}, 284-304 (1940).
\item $[23]$ F. M. Kuni, A. A. Melikhov, T. Yu. Novozhilova, I. A. Terentev, 
Theor. Math. Phys. {\bf 83}(2), 530-542 (1990); TMF (in Russian), {\bf 83}(2) 274-290 (1990).
\item $[24]$ Kuni, F.M.; Melikhov, A.A. 
{\em Theor. Math. Phys.} {\bf 1989}, {\it 81(2)}, 1182-1194 (1989); 
TMF (in Russian) {\bf 81}(2), 247-262 (1989). 
\item $[25]$ R. G. Horn and C. R. Johnson, {\it Matrix Analysis} 
(Cambridge University Press, Cambridge, 2013). 

\end{list}

\end{document}


\bibliographystyle{plain}
\renewcommand{\baselinestretch}{1}
\maketitle
\renewcommand{\baselinestretch}{2}
\normalsize
\renewcommand{\baselinestretch}{2}

\newcounter{s2e}
\newcommand{\ale}{\setcounter{s2e}{\value{equation}}%
\stepcounter{s2e}\setcounter{equation}{0}\renewcommand{\theequation}{S\arabic{equation}}}
\ale 

\section{Derivation of equation (4)}

Denote the numbers of molecules of components $1$ (water), $2$ (hydrophilic organic), and $3$ (hydrophobic
organic) in an AHHO aerosol by $\nu_1,\nu_2$, and $\nu_3$, respectively, and the number of radicals \\
RO\hspace{-0.01cm}{\rad} (component 4) in the aerosol by $\nu_4$; for the sake of simplicity and uniformity,
radicals RO\hspace{-0.01cm}{\rad} will be sometimes referred to as ``molecules of component $4$".  Since
components 2 and 4 are mostly hydrophilic, they are assumed to be distributed more or less uniformly within
the aerosol, while molecules of component 3 will be located mostly on the aerosol surface (although their
dissolution within the aerosol core can not be excluded). 

Choosing $\nu_1,\nu_2,\nu_3$, and $\nu_4$ as the independent  variables of state of a single aerosol, consider
an ensemble of AHHO aerosols in the air and  denote their distribution function with   respect to
$\nu_1,\nu_2,\nu_3$, and $\nu_4$ at time $t$ by $g(\nu_1,\nu_2,\nu_3,\nu_4,t)$.  Let us construct a discrete
balance equation governing the temporal evolution of $g(\nu_1,\nu_2,\nu_3,\nu_4,t)$. Depending on the
convenience, any function $f$ of variables  $\nu_1,\nu_2,\nu_3,\nu_4$ can be denoted either
$f(\nu_1,\nu_2,\nu_3,\nu_4)$ or  $f(\{\nu\})$ or $f(\nu_i,\widetilde{\nu_i})$, where $\{\nu\}$ would denote
either the aerosol itself or, as a function argument, all four independent variables of state of the aerosol,
$\nu_1,\nu_2,\nu_3,\nu_4$, whereas the ``complementary" variable  $\widetilde{\nu_i}$ would represents only
three of them,  the ``excluded" variable being $\nu_i$ (the total number of molecules in the aerosol will be
denoted by $\nu=\nu_1+..+\nu_4$).  In this notation, e.g.,
$g(\{\nu\},t)=g(\nu_i,\widetilde{\nu_i},t)=g(\nu_1,\nu_2,\nu_3,\nu_4,t)$

As usual in CNT, let us first assume that the metastability of vapor mixture in the air is  created
instantaneously and does not change during the whole nucleation process; this assumption  can be subsequently
removed and the kinetic theory properly modified for application to more complicated (and more realistic)
environmental conditions.  The temperature $T$ of the air parcel and the number densities of non-condensible 
gas molecules are also fixed.   Aerosols during the nucleation stage are so small that  the characteristic
times of their internal relaxation processes are very small in comparison with the time between two of its
successive interactions with the air, and the interactions  themselves take place under a free-molecular
regime. Thus, one can assume that the aerosol attains its internal thermodynamical equilibrium  before each
successive interaction with air, so that  the aerosol temperature can be also assumed to be equal to $T$.

\par The material exchange between an aerosol and air occurs via the following elementary interactions:\\ 
(a123) absorption of a molecule of component $1$ or $2$ or $3$ from the air into the aerosol 
$\{\nu\}$ with the rate $W^+_i=W^+_i(\{\nu\})$;\\ 
(e123) emission of a molecule of component $1$ or $2$ or $3$ from the aerosol $\{\nu\}$  into the air with the
rate $W^-_i=W^-_i(\{\nu\})$;\\ 
(f4) production of a ``molecule" of component 4 (radical RO\hspace{-0.01cm}{\rad}) via the 
forward sequence of heterogeneous chemical reactions (1)-(3) on the surface of aerosol 
$\{\nu\}$, with the aggregate rate $W^+_4=W^+_4(\{\nu\})$;\\
(b4) destruction of a ``molecule" of component 4 (radical RO\hspace{-0.01cm}{\rad}) via the 
backward sequence of heterogeneous chemical reactions (1)-(3) on the surface of aerosol 
$\{\nu\}$, with the aggregate rate $W^-_4=W^-_4(\{\nu\})$;
 
Therefore, the temporal evolution of the distribution
function $g(\nu_1,\nu_2,\nu_3,\nu_4,t)$ can be described by the discrete balance equation 
\beq \frac{\partial g(\nu_1,\nu_2,\nu_3,\nu_4,t))}{\partial t}=D_1+D_2+D_3+D_4, \eeq
where 
\be D_1= W^+_1(\nu_1-1,\nu_2,\nu_3,\nu_4)g(\nu_1-1,\nu_2,\nu_3,\nu_4,t)- 
W^+_1(\nu_1,\nu_2,\nu_3,\nu_4)g(\nu_1,\nu_2,\nu_3,\nu_4,t) + \nonumber \\
+ W_1^-(\nu_1+1,\nu_2,\nu_3,\nu_4)g(\nu_1+1,\nu_2,\nu_3,\nu_4,t)- 
 W_1^-(\nu_1,\nu_2,\nu_3,\nu_4)g(\nu_1,\nu_2,\nu_3,\nu_4,t),\ee
\be D_2= W^+_2(\nu_1,\nu_2-1,\nu_3,\nu_4)g(\nu_1,\nu_2-1,\nu_3,\nu_4,t)- 
W^+_2(\nu_1,\nu_2,\nu_3,\nu_4)g(\nu_1,\nu_2,\nu_3,\nu_4,t) + \nonumber \\+ 
W_2^-(\nu_1,\nu_2+1,\nu_3,\nu_4)g(\nu_1,\nu_2+1,\nu_3,\nu_4,t)- 
W_2^-(\nu_1,\nu_2,\nu_3,\nu_4)g(\nu_1,\nu_2,\nu_3,\nu_4,t), \ee
\be D_3= W^+_3(\nu_1,\nu_2,\nu_3-1)g(\nu_1,\nu_2,\nu_3-1,\nu_4,t)- 
W^+_3(\nu_1,\nu_2,\nu_3,\nu_4)g(\nu_1,\nu_2,\nu_3,\nu_4,t) + \nonumber \\+ 
W_3^-(\nu_1,\nu_2,\nu_3+1,\nu_4)g(\nu_1,\nu_2,\nu_3+1,\nu_4,t)- 
W_3^-(\nu_1,\nu_2,\nu_3,\nu_4)g(\nu_1,\nu_2,\nu_3,\nu_4,t), \ee
\be D_4= W^+_4(\nu_1,\nu_2+1,\nu_3,\nu_4-1)g(\nu_1,\nu_2+1,\nu_3,\nu_4-1,t)- 
W^+_4(\nu_1,\nu_2,\nu_3,\nu_4)g(\nu_1,\nu_2,\nu_3,\nu_4,t) + \nonumber \\+ 
W_4^-(\nu_1,\nu_2-1,\nu_3,\nu_4+1)g(\nu_1,\nu_2-1,\nu_3,\nu_4+1,t)- 
W_4^-(\nu_1,\nu_2,\nu_3,\nu_4)g(\nu_1,\nu_2,\nu_3,\nu_4,t), \ee
Note that  these equations assume the evolution of aerosols to occur through the absorption from and emission
into the vapor of single molecules of components $1,2$,  and $3$  (i.e., multimer absorption  and emission are
neglected),  as well as through the single sequences (1)-(3) of forward and backward reactions whereby a
radical  RO\hspace{-0.01cm}{\rad} is either formed or destroyed. 

The terms $D_1,D_2$, and $D_3$ on the RHS of eq.(S1) represent the contributions to  $\partial
g(\{\nu\},t)/\partial t$  from the material exchange events of type (a123) and (b123), whereas the term $D_4$ 
represents the contributions to $\partial g(\{\nu\},t)/\partial t$ from the elementary events of type (f4) and
(b4). Furthermore, on the RHS of the each of eqs.(S2)-(S4) the first two terms represent the contributions to
$\partial g(\{\nu\},t)/\partial t$ from the absorption events (a123) of  air molecules by aerosols, whereas
the third and forth terms therein are due to the emission of molecules from aerosols into the air. On the RHS
of eq.(S5), the first two terms represent the contributions to $\partial g(\{\nu\},t)/\partial t$ from the
forward sequences (1)-(3) of chemical reactions on aerosols (whereby radicals RO\hspace{-0.01cm}{\rad} are
produced), whereas the third and fourth terms therein are due to the backward sequences (1)-(3) (whereby
radicals RO\hspace{-0.01cm}{\rad} are destroyed). As clear from eq.(S5) and in consistency with the sequence
of chemical reactions (1)-(3), the change of the aerosol distribution due to the variable $\nu_4$ is always
accompanied by its change due to the variable $\nu_2$, while the latter can also change independently due to
the direct material exchange between aerosols and air.

In the space of variables $\nu_1,..,\nu_4$, let us introduce the flux of aerosols along the axis 
$\nu_i\;\;(i=1,2,3)$ at time moment $t$ as
\beq J_i(\{\nu\},t)= W^+_i(\nu_i-1,\widetilde{\nu_i})g(\nu_i-1,\widetilde{\nu_i},t)-
W_i^-(\{\nu\})g(\{\nu\},t)\;\;\;\;\;\;\;(i=1,2,3),\eeq
and the flux of aerosols along the axis $\nu_4$ as 
\beq J_4(\{\nu\},t)=W_4^+(\nu_1,\nu_2+1,\nu_3,\nu_4-1)g(\nu_1,\nu_2+1,\nu_3,\nu_4-1)-
W_4^-(\{\nu\})g(\{\nu\},t).\eeq 
Taking these definitions into account, one can rewrite equations (S1)-(S5) as 
\beq \frac{\partial g(\{\nu\},t))}{\partial t}=
-\sum_{i=1}^{3}[J_i(\nu_i+1,\widetilde{\nu_i},t)-J_i(\{\nu\},t)]-[J_4(\nu_1,\nu_2-1,\nu_3,\nu_4+1,t)-
J_4(\{\nu\},t)],
\eeq
which is equation (4) in the main text.

\section{Derivation of equation (7)}

In CNT, the expressions for the absorption rates $W_i^+(\nu_1,\nu_2,\nu_3,\nu_4)\;\;(i=1,2,3)$ are provided by
the gas-kinetic theory, as 
\beq W_i^+(\{\nu\})=\frac1{4}\alpha_in_iv_{\mbox{\tiny T}i}S(\{\nu\})\;\;\;(i=1,2,3),\eeq
where $\alpha_i,\;n_i,\;v_{\mbox{\tiny T}i}\;\;\;(i=1,2,3)$ are the sticking coefficient, number density, and
mean thermal velocity, respectively, of molecules of component $i$, and $S(\{\nu\})$ is the surface area of
the aerosol.  The emission rates $W_i^-(\{\nu\})\;\;(i=1,2,3)$ are determined through the corresponding
$W_i^+(\{\nu\})$ on the basis of the principle of detailed balance [18,19].$^{}$  According to the requirement
of this principle, for the equilibrium distribution function $g_{\mbox{\tiny e}}(\{\nu\})$ not only the entire
RHS of eq.(S8) is equal to 0,  but every equilibrium flux $J_i^{\mbox{\tiny e}}\equiv J_i^{\mbox{\tiny
e}}(\{\nu\})$
must be equal to 0. 

Applying this principle to the flux $J_i^{\mbox{\tiny e}}\;\;(i=1,2,3)$ and taking into account that
$g_{\mbox{\tiny e}}(\{\nu\})=N_f\exp[-F(\{\nu\})]$, where $N_f$ is the normalization factor and $F(\{\nu\})$
is the free energy of formation of an aerosol $\{\nu\}$ in thermal units $k_BT$,  we have 
\beq  
W_i^-(\{\nu\})=W^+_i(\nu_i-1,\widetilde{\nu_i})\exp\left[-\left(F(\nu_i-1,\widetilde{\nu_i})-
F(\{\nu\})\right)\right].
\eeq
In the near-critical region (i.e., vicinity of the ``saddle point" of the 5D free-energy surface determined by
the function $F=F(\{\nu\})$), aerosols are large enough, $\nu_i\gg 1\;\;(i=1,..,4)$, one can use conventional
for CNT approximations 
\beq W^+_i(\nu_i-1,\widetilde{\nu_i})\approx W^+_i(\{\nu\}),\;\;\;
F(\nu_i-1,\widetilde{\nu_i})\approx F(\{\nu\})-F'_i(\{\nu\}).\eeq
$F'_i(\{\nu\})=\partial F(\{\nu\})/\partial \nu_i\;\;\;(i=1,..,4)$  
Expanding the exponential in eq.(S10) in Taylor series and neglecting terms of the order 
of $1/\nu^2$ and smaller therein, we obtain 
\beq W_i^-(\{\nu\})\approx W^+_i(\{\nu\})[1+F'_i(\{\nu\})]\;\;\;(i=1,2,3).\eeq 
Note that this approximation is valid only for large enough aerosols $\nu\gg 1$ in the 
near-critical region.

Applying the principle of detailed balance to the flux $J_4^{\mbox{\tiny e}}$ and using similar
considerations, one can obtain  one can obtain an interesting {\em approximate}  relationship between the
aggregate rates of forward and backward sequence of heterogeneous chemical reactions  (1)-(3):
\beq W_4^-(\{\nu\})\approx W_4^+(\{\nu\})[1-(F'_2(\{\nu\})-F'_4(\{\nu\}))],\eeq
which is eq.(7) in the main text.   Clearly, this relationship between aggregate forward and backward reaction
rates of sequence (1)-(3) can be also used to obtain  a relationship between the corresponding reaction rate
constants; such a relationship  would depend on whether the heterogeneous reactions in sequence (1)-(3) are of
the first or second order.$^{}$

\section{Derivation of equation (8)}

In order to obtain eq.(8) from eq.(4), let us first expand in Taylor series (in deviations of 
$\nu_i\pm 1$ from $\nu_i$) the fluxes $J_i\;\;(i=1,..,4)$ on the RHS of eq.(4): 
\beq 
J_i(\nu_i+1,\widetilde{\nu_i},t)\simeq J_i(\{\nu\},t)+
\frac{\partial  J_i(\{\nu\},t)}{\partial \nu_i}\;\;\;\;(i=1,2,3),
\eeq
\beq 
J_4(\nu_1,\nu_2-1,\nu_3,\nu_4+1,t)\simeq J_4(\{\nu\},t)-\frac{\partial  J_4(\{\nu\},t)}
{\partial \nu_2}+\frac{\partial  J_4(\{\nu\},t)}{\partial \nu_4},
\eeq 
Here, it is taken into account that nucleating aerosols (in the near-critical region) are  large enough so
that $\nu_i\gg 1\;\;(i=1,..,4)$ and the terms with the second and higher order derivatives with respect to
$\nu_i\;\;(i=1,..,4)$ in the Taylor series expansions can be neglected, as usual in CNT [18,19,21].

Next, with the same degree of accuracy, one can obtain the Taylor series expansions (in deviations of 
$\nu_i\pm 1$ from $\nu_i$) of the distribution functions $g(\{\nu\},t)$ on the RHSs of eqs.(S6),(S7):
\beq
g(\nu_i-1,\widetilde{\nu_i},t)\simeq g(\{\nu\},t)-\frac{\partial  g(\{\nu\},t)}{\partial \nu_i}\;\;\;\;(i=1,2,3), 
\eeq
\beq
g(\nu_1,\nu_2+1,\nu_3,\nu_4-1,t)\simeq g(\{\nu\},t)+\frac{\partial  g(\{\nu\},t)}{\partial \nu_2}-
\frac{\partial  g(\{\nu\},t)}{\partial \nu_4}.
\eeq

Finally, according to the definition of the near-critical region, the width of the region is much smaller than
the size of the critical aerosol, i.e.,  $(\nu_i-\nu_{ic})/\nu_c\ll 1$ in the near-critical region. Therefore,
according to eq.(S9), one can assume that with a high degree of accuracy in that region $W_i^+(\{\nu\})\approx
W_{ic}^+$ (subscript ``c" marks quantity for the nucleus (critical aerosol). Taking this into account,
substituting eqs.(S14)-(S17) into the RHS of eq.(S8) and combining together appropriate terms, one can obtain 
\be \frac{\partial g(\{\nu\},t) }{\partial t}&=&\sum_{i=1}^{4}
W_{ic}^+\frac{\partial }{\partial \nu_i}
\left( F'_i(\{\nu\})+\frac{\partial }{\partial \nu_i} \right)g(\{\nu\},t)) \nonumber \\
&&+\;W_{4c}^+
\left[ \frac{\partial }{\partial \nu_2}
\left(F'_2(\{\nu\})+\frac{\partial }{\partial \nu_2}\right)\right.\nonumber\\
&&\left.\hspace{9mm}-\;
\frac{\partial }{\partial \nu_2}
\left(F'_4(\{\nu\})+\frac{\partial }{\partial \nu_4}\right)-
\frac{\partial }{\partial \nu_4}\left(F'_2(\{\nu\})+\frac{\partial }{\partial \nu_2}\right)
\right]g(\{\nu\},t)), 
\ee 
which is the Fokker-Planck equation (8) of the main text.